\documentclass[11pt,onecolumn,superscriptaddress,showpacs,nofootinbib,notitlepage]{article}
\usepackage{amsmath}
\usepackage{latexsym}
\usepackage{amssymb}
\usepackage{graphicx}
\usepackage[colorlinks=true, citecolor=blue, urlcolor=blue]{hyperref}
\usepackage{float}
\usepackage{amsfonts}
\usepackage{textcomp}
\usepackage{authblk}
\title{The emergence of Cooperation in the thermodynamic limit}
\author[1]{Colin Benjamin}
\author[1]{Shubhayan Sarkar}
\affil[1]{School of Physical Sciences, National Institute of Science Education and Research, HBNI, Jatni- 752050, India}
\begin{document}
\maketitle
\begin{abstract}
Predicting how cooperative behavior arises in the thermodynamic limit is one of the outstanding problems in evolutionary game theory. For two player games, cooperation is seldom the Nash equilibrium. However, in the thermodynamic limit cooperation is the natural recourse regardless of whether we are dealing with humans or animals. In this work, we use the analogy with the Ising model to predict how cooperation arises in the thermodynamic limit.
\end{abstract}
\newpage
\section{Introduction}
The solution to any game theoretic problem involves finding an equilibrium strategy known as the Nash equilibrium, whence deviating from this strategy brings in more loss to the player. An introduction to Nash equilibrium for two player two strategy games can be found in Ref.~\cite{4}. Finding the Nash equilibrium analytically for a two player game isn't difficult, however, many a situation arises wherein one needs to go beyond two players. To analyze such situations in a game theoretic setting one needs to go beyond two players to a game with infinite number of players, i.e., the thermodynamic limit. A particularly interesting problem arises in the context of evolution, where we see that cooperation arises even when defection is the preferred choice of individuals, see Ref.~\cite{5}. Game theory is inextricably linked with artificial intelligence. The problem of many or infinite number of agents collaborating is an outstanding problem not just in  game theory but in artificial intelligence too\cite{ai}. Cooperation in short term might not seem beneficial however, in the long run the population which cooperates survives. This scenario is generally tackled numerically and dynamically using replicator equations as has been shown in Refs.~\cite{12,nowak}. Further, in Ref.~\cite{13} the authors use the replicator equations to numerically calculate how the fraction of cooperators and defectors evolve in time when playing the Prisoner's dilemma game. The evolution of cooperators and defectors is not just restricted to Prisoner's dilemma, this has also been observed in context of social dilemmas like the Vaccination game, in Refs.~\cite{21,22}. The Vaccination game, which is a tool to monitor public health, is essentially a variation of the Prisoner's dilemma wherein instead of coooperators and defectors one has immunizers and anti-immunizers. In this paper, on the other hand, we use statistical mechanics tools to analytically predict whether and how cooperation emerges in the thermodynamic limit and this is one of the main attractions of our work. An approach has also been suggested by  Szabo and Borsos by constructing potentials corresponding to particular games and then comparing with models borrowed from statistical mechanics.  However, our approach is analytical and mathematically much more simpler than the method of Szabo and Borsos, see Ref.~\cite{19}. This is an interesting topic for evolutionary game theory and useful application of artificial intelligence, as any ecosystem tends to evolve towards an equilibrium with a large number of living beings. We show that 1D statistical models can correctly be applied to such social dilemmas.

Using the analogy with the 1D Ising model, we try to understand the equilibrium strategy in a population and predict how cooperative behavior emerges in the thermodynamic limit. We model a situation similar to 1D Ising model, where the sites are replaced by players and spin up or spin down correspond to the strategies $s_1$ or $s_2$ adopted by the players. Magnetization in Ising model is defined as the difference in the number of spin up and spin down particles. Similarly, Magnetization in game theory can be defined as the difference in the number of players choosing strategy $s_1$ or $s_2$. We first relate the Ising model to the payoff's in game theory and then apply this method first to Prisoner's dilemma and then to game of Chicken (variant of Hawk-Dove game). There have been earlier attempts to use the 1D Ising model to find the equilibrium strategy in the thermodynamic limit, see Ref.~\cite{5}. We find that there are some unphysical implications of the results of Ref.~\cite{5}. This paper is organized as follows-we next discuss how to connect the 1D Ising model to the payoffs of game theory by extending the analogy of Ref.~\cite{1} to the thermodynamic limit, then  we calculate the game magnetization which gives the Nash equilibrium strategy for Prisoner's dilemma in the thermodynamic limit. We observe how cooperators arise in Prisoner's dilemma in the thermodynamic limit even when defection is the Nash equilibrium. Further, we deal with the problems associated with the model of Ref.~\cite{5} in brief. Later we do a similar analysis for the game of Chicken which has no unique pure strategy Nash equilibrium in the two player case. We find how in the thermodynamic limit majority of cooperators can emerge. We end with the conclusions.

\section{1D Ising model and game theory}
\begin{figure}[h!]
\begin{center}
\includegraphics[width=\linewidth]{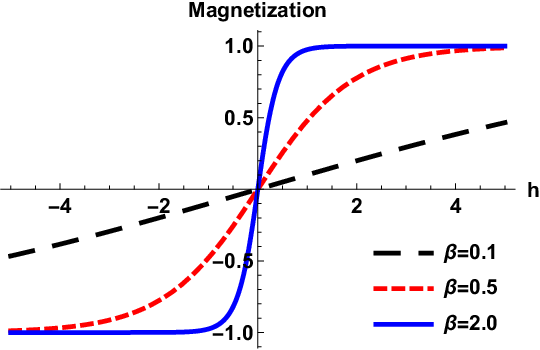}
  \caption{Variation of Magnetization with the external magnetic field $h$ for $J=.05$.}
  \label{fig1:}
  \end{center}
\end{figure}

There have been several works which have found it useful to explain the behavior in social dilemmas by looking at 1D models like the 1D Ising model,  see  Ref.~\cite{14} wherein the equivalence between the voter model and 1D Ising model has been brought out. Further, the 1D Sznajd model has been used as a template to explain spreading of opinion in a society, see Ref.~\cite{15}. These are just two examples which have used 1D Ising models to model social dilemmas, many others also do exist, see Ref.~\cite{14} for more examples.
The most well known among such models is the 1D Ising model\cite{6} which consists of spins that can be in either of the two states $+1$ ($\uparrow$)  or $-1$ ($\downarrow$). The spins are arranged in a line, and can only interact with their nearest neighbors. The Hamiltonian of such a system can be written as-
\begin{equation}\label{eq10}
H=-J\sum^N_{i=1}\sigma_i\sigma_{i+1}-h\sum^N_{i=1}\sigma_i,
\end{equation}
where $J$ is the coupling between the spins, $h$ is the external magnetic field and $\sigma$'s denote the spin. The partition function corresponding to the Hamiltonian~(\ref{eq10}) is
\begin{equation}\label{eq1}
Z=\sum_{\sigma_1}...\sum_{\sigma_N}e^{\beta(J\sum^N_{i=1}\sigma_i\sigma_{i+1}+h/2\sum^N_{i=1}(\sigma_i+\sigma_{i+1}))},
\end{equation}
$\beta=\frac{1}{k_{B} T}$, with $k_B$ being Boltzmann's constant. $\sigma_i$ denotes either the spin up $(+1)$ or spin down $(-1)$. In order to carry out the spin sum, we define a matrix $T$ with elements as follows,
\begin{eqnarray}
<\sigma|T|\sigma'>&=& e^{\beta(J\sigma\sigma'+h/2(\sigma+\sigma'))}\nonumber.
\end{eqnarray}
Using the transfer matrix $T$ and carrying out the spin sum via the completeness relation, the partition function from Eq.~(\ref{eq1}) in the large $N$ limit can be written as-
\begin{equation}
Z=e^{N\beta J}(\cosh(\beta h)\pm \sqrt{\sinh^2(\beta h)+e^{-4\beta J}})^N.
\end{equation}
Since the Free energy $F=-k_{B}T \ln Z$, the magnetization is
\begin{equation}\label{eq8}
m=-\frac{dF}{dh}=\frac{\sinh(\beta h)}{\sqrt{\sinh^2(\beta h)+e^{-4\beta J}}}.
\end{equation}
In Fig.~\ref{fig1:}, we plot the magnetization versus external magnetic field $h$ for different values of inverse temperature($\beta$).
In Ref.~[\cite{1}], it has been shown that a one-to-one correspondence can be made between 1D Ising model Hamiltonian and the payoff matrix for a particular game. We first look at a general payoff matrix for two player game and try to understand the methodology,
\begin{equation}\label{eq9}
U=\left(\begin{array}{c|cc}& s_1 & s_2 \\\hline s_1 & a,a' & b,b' \\  s_2 & c,c' & d,d'\end{array}\right),
\end{equation} 
where $U(s_i,s_j)$ is the payoff function with $a, b, c, d$ as the payoffs for row player and $a', b', c', d'$ are the payoffs for column player, $s_1$ and $s_2$ denote the strategies adopted by the two players. The Nash equilibrium as defined before is the strategy, deviating from which brings loss to the players. For the payoff's defined in Eq.~(\ref{eq9}), the Nash equilibrium, in case one takes the condition of a symmetric game, i.e., $b'=c, c=b'$ and $a=a', b=b'$ with $c<d<a<b$, is the strategy $(s_{2},s_{2})$. To make a one-to-one correspondence of the game payoff's with Ising model we need a set of transformations to the payoff's which will achieve that.  The transformations of payoffs of the players are as follows:
\begin{equation}\label{eq2}
U=\left(\begin{array}{c|cc} & s_1 & s_2 \\\hline s_1 & a+\lambda,a'+\lambda' & b+\mu, b'+\lambda'\\  s_2 & c+\lambda,c'+\mu' & d+\mu, d'+\mu'\end{array}\right).
\end{equation} 
 We choose the transformations as $\lambda=-\frac{a+c}{2},\lambda'=-\frac{a'+b'}{2}$ and $\mu=-\frac{b+d}{2},\mu'=-\frac{c'+d'}{2}$. Under such a transformation the Nash equilibrium doesn't change (see Supplementary material section 1.2 for more details where we prove this result using fixed point analysis, see also Ref.~\cite{11}).  Since Ising model Hamiltonian considered above \eqref{eq1}, assumes that the coupling $J$ is symmetric, thus to model the game \eqref{eq2} using Ising model we must consider only symmetric games, i.e., $a=a',\ b=c',\ c=b'$ and $d=d'$. 
After imposing the above conditions the transformed payoff matrix from \eqref{eq2} becomes-
\begin{equation}\label{eq12}
U=\left(\begin{array}{c|cc} & s_1 & s_2 \\\hline s_1 & \frac{a-c}{2},\frac{a-c}{2} & \frac{b-d}{2},\frac{c-a}{2} \\  s_2 & \frac{c-a}{2},\frac{b-d}{2} & \frac{d-b}{2},\frac{d-b}{2}\end{array}\right).
\end{equation} 

To calculate the Nash equilibrium of a generalised two player game in the thermodynamic limit, we have to relate the transformed payoff matrix of the classical game as in Eq.~(\ref{eq12}) to the Ising model Hamiltonian with two spins. When $N=2$, the Hamiltonian Eq.~\eqref{eq10} can be written as-
\begin{equation}\label{eq4}
H=-J(\sigma_1\sigma_{2}+\sigma_2\sigma_{1})-h(\sigma_1+\sigma_{2})
\end{equation}
So the individual energies of the spins 1 and 2 can be written as: 
\begin{equation}\label{eq21}
E_1=-J\sigma_1\sigma_{2}-h\sigma_1,
E_2=-J\sigma_2\sigma_{1}-h\sigma_{2}.
\end{equation}
It is to be noted that equilibrium in Ising model corresponds to minimizing the energies of spins. Now for symmetric coupling as in Eq.'s~(\ref{eq1},\ref{eq4}) minimizing Hamiltonian $H$ with respect to spins $\sigma_1, \sigma_2$ is same as maximizing $-H$ with respect to $\sigma_1, \sigma_2$. In game theory, on the other hand players search for the Nash equilibrium with maximum payoffs. This implies maximizing the payoff function $U$ in Eqs~(\ref{eq9}-\ref{eq12}) with respect to strategies $s_i,s_j$ which for the two player Ising model is equivalent to maximizing $-E_i$, see Eq.~\eqref{eq21} with respect to spins $\sigma_i,\sigma_j$. Thus, the Ising game matrix can be written as-
\begin{equation}\label{eq13-}
E=\left(\begin{array}{c|cc}  & \sigma_2=+1 & \sigma_2=-1  \\\hline \sigma_1=+1 & J+h,J+h & -J+h,J-h \\\sigma_1=-1 & J-h,-J+h & -J-h,-J-h \end{array}\right).
\end{equation}
Comparing the matrix elements of the transformed payoff matrix- Eq.~(\ref{eq12}) to the Ising game matrix Eq.~\eqref{eq13-}, we get the relation between parameters of Ising model ($J$ and $h$) and the payoffs of two player game as- 
\begin{equation}
J=\frac{a-c+d-b}{4},\ h=\frac{a-c+b-d}{4}.
\end{equation} 
Substituting $J$ and $h$ in terms of payoff's in the equation for magnetization\eqref{eq8}, gives us the game magnetization ($m_{g}$), defined as the fraction of player's choosing strategy $s_1$ over $s_2$ in the thermodynamic limit-
\begin{equation}\label{eq8g}
m_{g}=-\frac{df}{dh}=\frac{\sinh(\beta h)}{\sqrt{\sinh^2(\beta h)+e^{-4\beta J}}}=\frac{\sinh(\beta \frac{a-c+b-d}{4})}{\sqrt{\sinh^2(\beta \frac{a-c+b-d}{4})+e^{-4\beta \frac{a-c+d-b}{4}}}}.
\end{equation}
This completes the connection of the payoffs from a two player game to Ising model relating spins in the thermodynamic limit. $\beta$ in Ising model is the inverse temperature. Decreasing  $\beta$ or increasing temperature leads to randomness in spin orientation. Thus, as $\beta\rightarrow 0$ then game magnetization vanishes, see \eqref{eq8g}, due to increase in randomness in the strategic choices of the players. In the following sections we will apply this to some famous two player games so as to analyze them in the thermodynamic limit. Note that this approach can't be compared to Nowak's approach as in contrast to replicator equations, the above suggested approach is not dynamical but completely analytical. However, in the next section we qualitatively compare our results with the conclusions presented in Refs.~\cite{12,13}. 
\section{Prisoner's dilemma}
In this game, the police are questioning two suspects in separate cells. Each has two choices: to cooperate with each other and not confess the crime (C), or defect to the police and betray each other(D). We construct the Prisoner's dilemma payoff matrix by taking the matrix elements from Eq.~\eqref{eq9} as $a=r$, $d=p$, $b=s$ and $c=t$, with $t>r>p>s$ where $r$ is the reward, $t$ is the temptation, $s$ is the sucker's payoff and $p$ is the punishment. Thus, the payoff matrix is- 
\begin{equation}\label{eq13}
U=\left(\begin{array}{c|cc} & C & D \\\hline C & r,r & s,t \\  D & t,s & p,p\end{array}\right).
\end{equation} 
The values in the payoff matrix can be explained as follows- reward $r$ means 1 year in jail while punishment $p$ means 10 years in jail, sucker's payoff $s$ represents a life sentence while temptation $t$ implies no jail time. Independent of the other suspects choice, one can improve his own position by defecting. Therefore, the Nash equilibrium in this case is to defect. Following on from the calculations for the general two player game as in Eq.~\eqref{eq8} and Eq.~\eqref{eq8g} as applied to Prisoners dilemma game matrix Eq.~\eqref{eq13}, we get-
$J=\frac{r-t+p-s}{4}$ and $h=\frac{r+s-t-p}{4}$. From Ising model, the game magnetization($m_g$) in the thermodynamic limit Eq.~(\ref{eq8}) is-
\begin{equation}\label{eq14}
m_{g}=\frac{\sinh(\beta \frac{r+s-p-t}{4})}{\sqrt{\sinh^2(\beta \frac{r+s-t-p}{4})+e^{-\beta (r-t+p-s)}}}.
\end{equation} 
\begin{figure}[h!]
\begin{center}
\includegraphics[width=\linewidth]{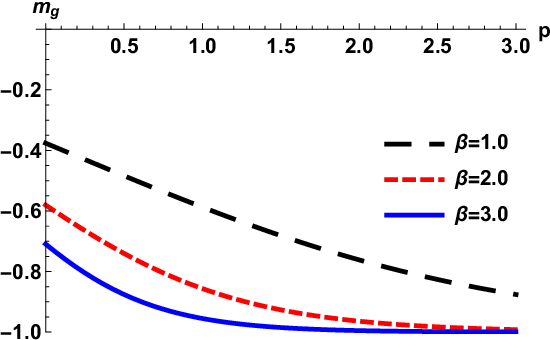}
  \caption{Variation of game Magnetization ($m_{g}$) with the punishment $p$ (defect payoff) for Prisoner's dilemma for $t=5$, $s=0$ and $r=3$. Lowering punishment from 1 to .5 increases the number of cooperators by $30~\%$.}
  \label{fig2:}
  \end{center}
\end{figure}
Plotting game magnetization as in Eq.~\eqref{eq14} as function of punishment, with the condition: $s<p<r$, we see in the thermodynamic limit the Nash equilibrium is always the defect strategy. A phase transition would occur only if $p<r+s-t$ which is not possible as the Prisoner's dilemma has the condition: $p>s$ and $t>r$. When $\beta$ decreases, game magnetization decreases, which implies that number of cooperators increases. As $\beta\rightarrow 0$, $m_{g}\rightarrow$ 0, implying equal number of cooperators and defectors. At finite and large $\beta$ as seen from Fig.~\ref{fig2:}, in the thermodynamic limit for $p>1.5$ almost all are defectors. However, in the range $0<p<1.5$ there is a decrease in the number of defectors so much so that around $p=.5$,  $25~\%$ of the population tend to cooperate for $\beta=1.0$. From the numerical calculations presented in Ref.~\cite{13}, we also see that the number of defectors always remains a majority and the number of cooperators in any particular generation increases if the reward increases which is similar to the results presented above. Further, in Ref.~\cite{12} it is shown that in the iterative Prisoner's dilemma with finite number of players cooperation can become Nash equilibrium if tit for tat scheme is allowed for some fraction of population but not the entire population. In Ref.~\cite{12}, the thermodynamic limit is not diretly dealt with however, they infer via natural selection that for the iterative Prisoner's Dilemma the Nash equilibrium would be everyone defecting. Contrary to this we show that even if defection is the Nash equilibrium in the thermodynamic limit there still exist a finite minority of cooperators which too increase as the reward increases. Thus, our results show that even in the thermodynamic limit of Prisoner's dilemma cooperation emerges. In the next section we approach this problem via the method proposed in Ref.~\cite{5} and unravel some deficiencies in the method of Ref.~\cite{5}. 
\subsection{Problems with the approach of Ref.~\cite{5}}
The connection between Ising model and game theory as shown above is not the only approach available. In Ref.~\cite{5} too, it has been shown that in the thermodynamic limit games can be modeled using 1D Ising model. However, when one analyses the Prisoner's dilemma game using the approach of Ref.~\cite{5}, the results are not compatible with the basic tenets of the game for some cases, as shown below.
\subsubsection{ When reward $r$ approaches temptation $b$:} We start with payoff matrix\eqref{eq17} used in Ref.~\cite{5}, to describe Prisoner's dilemma- 
\begin{equation}\label{eq17}
U=\left(\begin{array}{c|cc} & C & D \\\hline C & r,r & -c,b \\  D & b,-c & 0,0\end{array}\right).
\end{equation}
where $r=b-c$, with $b>r>0$ and $b>c>0$. Eq.~(\ref{eq17}) is the payoff matrix used in Ref.~\cite{5}. This is similar to Eq.~(\ref{eq13}) with payoffs for reward as $r$, temptation as $b$, sucker's payoff as $-c$ and punishment  as $0$ with the condition $r=b-c$. The game magnetization as derived in Ref.~\cite{5} is given as- 
\begin{equation}\label{eq16}
m_{g}= \frac{e^{-\beta r}-1}{(1+e^{-\beta r})}.
\end{equation}
This game magnetization is independent of temptation $b$ unlike that derived in Eq.~(\ref{eq14}) using our approach. Although in Ref.~[\cite{5}] it has been shown that for all values of reward $r$, the dominant choice is to defect but this is not true in the limiting case when $r$ approaches $b$. We analyze the same situation using the payoff matrix of the Prisoner's dilemma-
\begin{equation}\label{eqpd-pay-2}
U=\left(\begin{array}{c|cc} & C & D \\\hline C & b,b & 0,b \\  D & b,0 & 0,0\end{array}\right).
\end{equation}
In Eq.~\eqref{eqpd-pay-2}, we see an inconsistency, when reward $r$ equals the temptation $b$, there is no unique Nash equilibrium, i.e., both strategies cooperation and defection are equiprobable. The players can equally choose between cooperation and defection and hence game Magnetization $m_{g}$ should be $0$. However, from Ref.~[\cite{5}] the game magnetization~\eqref{eq16} is negative (see Fig.~\ref{fig4:} inset) which means that defect is the Nash equilibrium which is not correct. 
\subsubsection*{ The reward $r$ approaches $0$:} 
\begin{figure}[h!]
\begin{center}
\includegraphics[width=\linewidth]{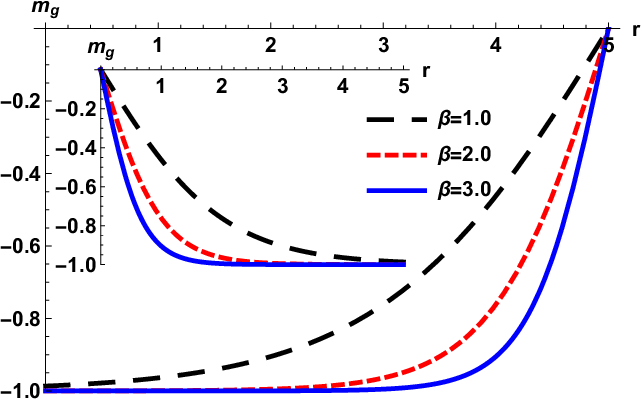}
  \caption{Variation of game magnetization $m_g$ with $r$ for Prisoner's dilemma when $b=5$. For all values of $b>r>0$ there is no phase transition. When $r=b=5$, the game magnetization is $0$ as expected. Also when $r=0$, game magnetization is negative which means defect is the Nash equilibrium. Inset: Variation of game magnetization with reward $r$ for Prisoner's dilemma as calculated in Ref.~[\cite{5}].  When $r=b$, (say $b=5$) game Magnetization $\rightarrow -1$ and when $r=0$ game magnetization vanishes. These results are not compatible with the definition of Prisoner's dilemma. }
  \label{fig4:}
  \end{center}
\end{figure}

Another situation where Ref.~[\cite{5}]'s results are negated is when $r=0$, $m_g$ tends to $0$ as in Eq.~(\ref{eq16}) implying equal number of cooperators and defectors. However, when we look at the payoff matrix Eq.~(\ref{eq17}) for $r=0$, we have-
\begin{equation}
U=\left(\begin{array}{c|cc} & C & D \\\hline C & 0,0 & -b,b \\  D & b,-b & 0,0\end{array}\right),
\end{equation} 
one can see defect (D,D) is still the Nash equilibrium. Using our approach, see the calculations as done in Eqs.~\eqref{eq9}-\eqref{eq8g} and using payoff matrix \eqref{eq17} we get the game magnetization as $m_{g}=\tanh(\beta \frac{r-b}{2})$ where we have substituted $t=b,s=-c,p=0$ with the condition $r=b-c$ in Eq.~(\ref{eq14}). In Fig.~\ref{fig4:} one sees $m_{g} \rightarrow -1$ as $r\rightarrow 0$, which is the correct result using our approach. In inset of Fig.~\ref{fig4:}, the game magnetization using the approach of Ref.~[\cite{5}] however tends to $0$ which is obviously incorrect.
From Fig.~\ref{fig4:} as reward $r$ approaches temptation $b$, game magnetization $m_{g}\rightarrow 0$. Further, when reward $r$ approaches $0$, game magnetization $m_{g}\rightarrow -1$. Our approach corrects the problems in Ref.~\cite{5} in the limiting cases when $r\rightarrow 0$ and $r \rightarrow b$. This is also elaborately dealt with in Ref.~\cite{18} along with the case of Public goods game with and without punishment. In the supplementary material accompanying this article we deal elaborately with the reasons behind the problems in approach of Ref.~\cite{5}.
 Next, we extend this approach to the game of Chicken.
\section*{ Game of Chicken} The name ``Chicken" has its origins in a game in which two teenagers drive their vehicles towards each other at high speeds\cite{4}. Each has two strategies: one is to swerve and the other is going straight. If one teenager swerves and the other drives straight, then the one who swerved will be called a "Chicken" or coward. ``Hawk$-$Dove" game, on the other hand refers to a situation in which players compete for a shared resource and can choose either mediate (Dove strategy) or fight for the resource (Hawk strategy). The parameterized payoff matrix from Eq.~(\ref{eq9}) by taking $a=-s,\ b=r,\ c=-r$ and $d=0$ for the game of Chicken is given by-
\begin{equation}\label{eq15}
U=\left(\begin{array}{c|cc} & straight & swerve \\\hline straight & -s,-s & r,-r \\  swerve & -r,r & 0,0\end{array}\right),
\end{equation} 
where $``r"$ denotes the reputation and $``s"$ denotes the cost of injury and $s>r>0$. If one teen swerves before the other, then the one who drives straight gains in reputation while the other loses reputation. However, if both drive straight, there is a crash, and both are injured. There are two pure strategy Nash equilibriums (straight, swerve) and (swerve, straight). Each gives a payoff of $r$ to one player and -$r$ to the other. There is another mixed strategy Nash equilibrium given by $(\sigma,\sigma)$, where [$\sigma$ = $p$.straight+$(1-p)$.swerve] where $p=\frac{r}{s}$ ($p$ is the probability to choose straight). In ``Hawk-dove" the reputation from game of ``Chicken" is replaced by the value of resource and the cost of injury doesn't change. Similar to game of ``Chicken", the Hawk-Dove game has two pure strategy Nash equilibrium: (Hawk, Dove) and (Dove, Hawk) and a mixed strategy Nash equilibrium ($\sigma,\sigma$): [$\sigma$ = $p$.Hawk +$(1-p)$.Dove]. Thus, from a game-theoretic point of view, ``Chicken" and ``Hawk$-$Dove" are identical. 
\begin{figure}[h]
\begin{center}
\includegraphics[width=\linewidth]{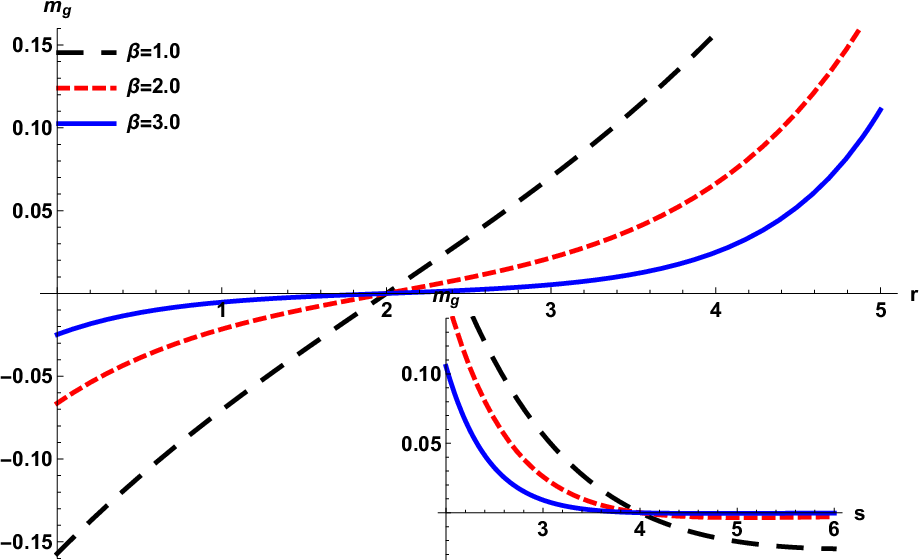}
  \caption{Variation of game magnetization with the reputation $r$ for the game of Chicken for $s=4$ for different values of the temperature. The phase transition occurs when $r=s/2=2$ as expected. Inset: Variation of game magnetization ($m_{g}$) with the cost of injury $s$ for game of Chicken when reputation $r=2$ for different values of the temperature. For lower cost of injury majority of the players choose straight or defection.}
  \label{fig3:}
  \end{center}
\end{figure}
We analyze the game of Chicken in the thermodynamic limit. Following on from the calculations for the general two player game as in Eq.~(4) and Eq.~(10) as applied to the game of Chicken payoff's in Eq.~(\ref{eq15}) we get $J=-\frac{s}{4}$ and $h=\frac{2r-s}{4}$. Thus, in the thermodynamic limit of the game of Chicken the game magnetization is-
\begin{equation}\label{eq18}
m_{g}=\frac{\sinh(\beta h)}{\sqrt{\sinh^2(\beta h)+e^{-4\beta J}}}=\frac{\sinh(\beta \frac{2r-s}{4})}{\sqrt{\sinh^2(\beta \frac{2r-s}{4})+e^{\beta s}}}.
\end{equation} 
From Eq.~(\ref{eq18}), the condition for change of sign in ``$m_{g}$" is given by-
\begin{equation}
\frac{\sinh(\beta \frac{2r-s}{4})}{\sqrt{\sinh^2(\beta \frac{2r-s}{4})+e^{\beta s}}}=0
\implies s=2r.
\end{equation} 
Plotting game magnetization $m_g$ as in Eq.~(\ref{eq18}), we see from Fig.~\ref{fig3:} that as the reputation $r$ increases more than $s/2$, more players choose straight as choosing swerve would bring in more loss to one's reputation. Similarly, when $r<s/2$ then players would rather choose to swerve and not get injured. Further, it should be noted from Eq.~(\ref{eq18}) that as the cost of injury increases (see inset of  Fig.~\ref{fig3:}), the game magnetization becomes more positive implying that more players choose to swerve or cooperate. 
\section*{Discussion and Conclusions}
Since game of Chicken and Hawk-Dove game are equivalent in game theory, it can be inferred from the above results that for Hawk-Dove game as the value of resource increases keeping the cost of injury constant, then more fraction of players choose the Hawk strategy (defect) or fight for the resource. Further, when the cost of injury increases then the players are reluctant to fight for the resource as getting injured is more expensive. Thus, larger fraction of players end up choosing Dove strategy(cooperate), i.e., sharing the resource when cost of injury is high. Contrary to the notion as in two player games that the players would always opt for the Nash equilibrium strategy, in the thermodynamic limit this is not true. In the thermodynamic limit our results show that a larger fraction of the players would choose the Nash equilibrium strategy but not every player. For example, when the temptation decreases in Prisoner's dilemma the fraction of cooperators increases even when the Nash equilibrium is to defect. A natural extension in the thermodynamic limit would be that every player would choose to defect as Nash equilibrium for the two player case is (D, D) however, there is a finite fraction of players who choose cooperation which increases as the temptation (t) decreases. Further, we see in game of Chicken that even if the reputation becomes high still there is a small fraction of players who choose to swerve(cooperate) and lose. This shows that in the thermodynamic limit cooperation does emerge even when defection would be the preferred choice of the individual players. Again, we see in Prisoner's dilemma that in thermodynamic limit slightly reducing  punishment below $r/3$ where $r$ is the reward increases the fraction of cooperators by a large amount even when the Nash equilibrium is to defect. Even in game of ``Chicken", when cost of injury is low the best choice for the players is to choose straight or defect. However, we find that still there exist a large fraction of players who choose to swerve or cooperate. In related works, see Refs.~\cite{10,sar-benj-physa}, we extend our model of predicting cooperative behavior in the thermodynamic limit to the quantum regime.
\section{Appendix}
In this supplementary material we first detail the reasons for error in the approach of Adami and Hintze, Ref.~[\cite{5}] and then we show how these errors are not seen in our approach. We also prove that the Nash equilibrium is invariant under transformations we use in our approach. 
\subsection {Reasons for error in approach of Ref.~[1]}
The main reason for the error in Ref.~\cite{5}, is that the maximization of the payoffs doesn't take into account the choice of the neighbouring players and occurs only over the payoffs of the individual player. Here, we find the payoffs for just the two player case of any general game using first the approach of Ref.~\cite{5} and then using our approach as well. We show that the approach of Ref.~\cite{5} gives incorrect average payoff implying that the maximization scheme used in Ref.~\cite{5} is not correct. In contrast our approach gives correct average payoff.
\subsubsection {Mapping Ising model and payoffs, approach of Ref.~[1]}
Let's consider a two spin system, and find the average payoff. The Hamiltonian for a $2$-spin system is defined using the approach and notation of Ref.~\cite{5} as- 
\begin{equation}
H=\sum_{m,n=0,1}U_{mn}P_m^{(1)}\otimes P_n^{(2)}+\sum_{m,n=0,1}U_{nm}P_m^{(2)}\otimes P_n^{(1)}
\label{H_app2}
\end{equation}
The above equation is Eq.~(2) of Ref.~\cite{5} for $N=2$. $U_{mn}$ represents the element of the payoff matrix $U$ for the $m^{th}$ row and $n^{th}$ column. The $P's$ represents projectors similar to the $\sigma's$ in the Ising model Hamiltonian (Eq.~(1) of our manuscript) and $P_0^{(1)}=|0\rangle\langle 0|$ and $P_1^{(1)}=|1\rangle\langle 1|$. 
$P_m^{(1)}$ defines the projector of site $(1)$ with $m$ either $0$ (meaning  $P_1^{(1)}= |0\rangle\langle0|$) or if at site $(1)$ again with $m$ being $1$ then  $P_1^{(1)}=|1\rangle\langle 1|$. Similarly $P_m^{(2)}$  defines the projector of site $(2)$ with $m$ taking values $0$ or $1$.  $m, n$ denote the indices for strategy which for $0$ denotes  cooperation while when $1$ denotes defection. 
Expanding the Hamiltonian Eq.~\eqref{H_app2} gives,
\begin{eqnarray}
H&=&U_{00}P_0^{(1)}\otimes P_0^{(2)}+ U_{01}P_0^{(1)}\otimes P_1^{(2)}+U_{10}P_1^{(1)}\otimes P_0^{(2)}+
U_{11}P_1^{(1)}\otimes P_1^{(2)}\nonumber\\&& +U_{00}P_0^{(2)}\otimes P_0^{(1)}+ U_{01}P_1^{(2)}\otimes P_0^{(1)}+ 
U_{10}P_0^{(2)}\otimes P_1^{(1)}+ U_{11}P_1^{(2)}\otimes P_1^{(1)}\nonumber\\
&=& 2U_{00}P_0^{(1)}\otimes P_0^{(2)}+(U_{01}+U_{10})P_0^{(1)}\otimes P_1^{(2)}+(U_{10}+U_{01})P_1^{(1)}\otimes P_0^{(2)}+2U_{11}P_1^{(1)}\otimes P_1^{(2)}.
\end{eqnarray}
The state of the two spin system can written as- $|x\rangle=|m_1m_2\rangle$. $|m_1\rangle  (|m_2\rangle)$ represent the state of the first (second) spin which can either be $|0\rangle$ or $|1\rangle$.  Thus, the partition function for the two spin system is-
\begin{eqnarray}
Z=Tr(e^{-\beta H})&=&\sum_{x}\langle x|e^{-\beta H}|x\rangle=\sum_{m_1,m_2=1,2}\langle m_1m_2|(1-\beta H+\frac{(\beta H)^2}{2!}...)|m_1m_2\rangle, \nonumber\\
&=& 1-\sum_{m_1,m_2=1,2}\beta\langle m_1m_2|H|m_1m_2\rangle +\sum_{m_1,m_2=1,2}\frac{\beta}{2!}\langle m_1m_2|H^2|m_1m_2\rangle+....
\label{Z_app2}
\end{eqnarray} 
Since $P_m^{(1)}=|m_1><m_1|$ and $P_m^{(2)}=|m_2><m_2|$, we have $H|m_1m_2>= (U_{m_1m_2}+U_{m_2m_1})|m_1m_2>$. Thus,
\begin{eqnarray}
\langle m_1m_2|H|m_1m_2\rangle&=& U_{m_1m_2}+U_{m_2m_1},\\
\mbox{ and } \langle m_1m_2|H^2|m_1m_2\rangle&=&( U_{m_1m_2}+U_{m_2m_1})\langle m_1m_2|H|m_1m_2\rangle=(U_{m_1m_2}+U_{m_2m_1})^2.
\end{eqnarray}
Therefore, from Eq.~\eqref{Z_app2} we have-
\begin{eqnarray}
Z&=&\sum_{m_1,m_2=1,2}[1-\beta(U_{m_1m_2}+U_{m_2m_1})+\frac{\beta^2}{2!}(U_{m_1m_2}+U_{m_2m_1})^2+....],\nonumber\\
\mbox{or, }Z&=&\sum_{m_1,m_2=1,2}e^{-\beta( U_{m_1m_2}+U_{m_2m_1})}= e^{-2\beta U_{00}}+e^{-\beta (U_{01}+U_{10})}+e^{-\beta (U_{10}+U_{01})}+e^{-2\beta U_{11}}.
\label{Z_app2-1}
\end{eqnarray}
The payoff matrix (see page 2, of Ref.~\cite{5}) obeys the condition- $U_{01}+U_{10}=U_{00}+U_{11}$. 
Now for a general payoff matrix for a symmetric game-
\begin{equation}
U=\left(\begin{array}{c|cc} & s_1 & s_2 \\\hline s_1 & a,a & b,c \\  s_2 & c,b & d,d\end{array}\right)\nonumber
\end{equation} 
wherein $U_{00}=a, U_{01}=b, U_{10}=c, U_{11}=d$, with the condition $a+d=b+c$, we have
\begin{eqnarray}
Z=(e^{-\beta U_{00}}+e^{-\beta U_{11}})^{2}=(e^{-\beta a}+e^{-\beta d})^{2}.\nonumber
\end{eqnarray}
The average payoff's can be found then by minimizing the Free energy in the no noise limit(zero temperature), i.e., $\beta\rightarrow \infty$, the average payoff using approach of Ref.~[\cite{5}] is then-
\begin{eqnarray}
\langle E\rangle=-\frac{\partial \ln Z}{\partial \beta}=-2\frac{\partial \ln(e^{-\beta d}+e^{-\beta a})}{\partial \beta}=2\frac{(de^{-2\beta d}+ae^{-\beta a})}{(e^{-\beta d}+e^{-\beta a})}
\end{eqnarray}
in the limit $\beta \rightarrow\infty$, the payoff for each player(dividing  $\langle E\rangle$ by $2$) becomes-
\begin{eqnarray}\label{eq29}
\lim_{\beta \rightarrow\infty}\langle E\rangle=\lim_{\beta \rightarrow\infty}\frac{(de^{-\beta d}+ae^{-\beta a})}{(e^{-\beta d}+e^{-\beta a})}
\end{eqnarray}
Let's consider two cases and check whether it gives the correct average payoff.
\begin{enumerate}
\item For $a>c$, $d<b$ a possibility exists such that  $a>c>b>d$. The Nash equilibrium for such a payoff matrix is the strategy $(s_1,s_1)$. However, from Eq.~\eqref{eq29}, we get-
\begin{eqnarray}
\lim_{\beta \rightarrow\infty}\langle E\rangle=\lim_{\beta \rightarrow\infty}\frac{(ae^{-\beta a}+de^{-\beta d})}{(e^{-\beta a}+e^{-\beta d})}=d
\end{eqnarray}
which is the payoff of strategy $(s_2,s_2)$. An incorrect conclusion.
\item  For $a<c$, $d>b$ a possibility exists such that  $d>b>c>a$. The Nash equilibrium for such a payoff matrix is the strategy $(s_2,s_2)$. However, from Eq.~\eqref{eq29}, we get-
\begin{eqnarray}
\lim_{\beta \rightarrow\infty}\langle E\rangle=\lim_{\beta \rightarrow\infty}\frac{(ae^{-\beta a}+de^{-\beta d})}{(e^{-\beta a}+e^{-\beta d})}=a
\end{eqnarray}
\end{enumerate}
which is the payoff of the strategy $(s_1,s_1)$. Thus, we have shown that even in the two player case, the approach of Ref.~[\cite{5}] gives incorrect average payoff. Now, lets check the average payoff calculation using our approach.
\subsubsection{Mapping Ising model and payoffs, using our approach}
 For two spin case the Hamiltonian from Eq.~(1) is,
\begin{equation}
H=-J(\sigma_1\sigma_{2}+\sigma_2\sigma_{1})-h/2(\sigma_1+\sigma_2).
\end{equation}
The partition function is given by-
\begin{equation}
Z=\sum_{\sigma_{1},\sigma_{2}} e^{\beta(J(\sigma_1\sigma_{2}+\sigma_2\sigma_{1})+h/2(\sigma_1+\sigma_2))}=e^{\beta(2J+h)}+2e^{-\beta(2J)}+e^{\beta(2J-h)}.
\end{equation}
As derived in manuscript Eq.~(10) of the main manuscript, for any general game $J=\frac{a-c+d-b}{4}$ and $h=\frac{a-c+b-d}{4}$. Putting the condition from Ref.~\cite{5}, $a+d=b+c$- we get $J=0$ and $h=\frac{a-c}{2}$. Thus, we have $Z=e^{\beta(h)}+2+e^{-\beta(h)}$.  The average payoff for 2 player case is then-
\begin{eqnarray}\label{eq19}
\langle E\rangle=-\frac{\partial \ln Z}{\partial \beta}=-\frac{\partial \ln(e^{\beta(h)}+2+e^{-\beta(h)}))}{\partial \beta}=\frac{-he^{\beta(h)}+he^{-\beta(h)}}{e^{\beta(h)}+2+e^{-\beta(h)}}
\end{eqnarray}
 However, unlike the approach of Ref.~\cite{5} which equates directly the energies to payoffs, in our approach while energies are minimized to find the equilibrium (point of lowest energy or ground state) the payoffs are maximized, meaning the payoffs should be taken as $-E$ as for game we intend to maximize the payoffs (whereas, Ising model aims to minimize the energies). The philosophy behind this is that in game theory players search for a Nash equilibrium with maximum payoff's. Thus one aims to maximize payoff's with respect to strategies ($s_{1},s_{2}$) which implies maximizing $-E$. Now, let's take two cases and see whether this gives the correct average payoff.
\begin{enumerate}
\item For $a>c$, $d<b$ a possibility exists such that  $a>c>b>d$. From game theory the Nash equilibrium is the strategy $s_1$. Using our approach from Eq.~\eqref{eq19}, for $a>c$  and as $h=\frac{a-c}{2}$ we get $h>0$. Thus, 
\begin{eqnarray}
\lim_{\beta \rightarrow\infty}-\langle E\rangle=\lim_{\beta \rightarrow\infty}\frac{he^{\beta(h)}-he^{-\beta(h)}}{e^{\beta(h)}+2+e^{-\beta(h)}}=h=\frac{a-c}{2}
\end{eqnarray}
which is the payoff of the strategy $(s_1,s_1)$ from the payoff matrices, see Eqs.~(5,6).
\item  For $a<c$, $d>b$ a possibility exists such that  $d>b>c>a$. From game theory the Nash equilibrium is the strategy $s_2$. Using our approach from Eq.~\eqref{eq19}, for $a<c$ and as $h=\frac{a-c}{2}$, we get $h<0$. Thus,
\begin{eqnarray}
\lim_{\beta \rightarrow\infty}-\langle E\rangle=\lim_{\beta \rightarrow\infty}\frac{he^{\beta(h)}-he^{-\beta(h)}}{e^{\beta(h)}+2+e^{-\beta(h)}}=-h=\frac{c-a}{2}=\frac{d-b}{2}.
\end{eqnarray}
which is the payoff of the strategy $(s_2,s_2)$ from the payoff matrices, see Eqs.~(5,6).
\end{enumerate}
Thus, our approach gives the correct payoffs corresponding to the Nash equilibrium strategies.

An essay by P Ralegankar, Understanding Emergence of Cooperation using tools from Thermodynamics, available at:\\ http://guava.physics.uiuc.edu/~nigel/courses/569/Essays$\_$Spring2018/Files/Ralegankar.pdf 
\\
also comes to similar conclusions.
 \subsection{Invariance of Nash equilibrium under transformations of Eq.~(6) in main manuscript}
Let's take the generalised game theory payoff matrix for two player games
\begin{equation}\label{eqa1}
E=\left(\begin{array}{c|cc}  & s_1 & s_2 \\\hline s_1 & a,a' & b,b' \\  s_2 & c,c' & d,d'\end{array}\right).
\end{equation} 
Transforming the elements of the payoff matrix by adding a factor $\lambda$ to column 1 and $\mu$ to column 2 for row player's payoffs and adding a factor $\lambda'$ to row 1 and $\mu'$ to row 2 for column player's payoffs   we get:
\begin{equation}\label{eqa2}
U=\left(\begin{array}{c|cc} & s_1 & s_2 \\\hline s_1 & a+\lambda,a'+\lambda' & b+\mu, b'+\lambda'\\  s_2 & c+\lambda,c'+\mu' & d+\mu, d'+\mu'\end{array}\right).
\end{equation} 
Herein, we give a simple proof that under such transformations the Nash equilibrium remains invariant using fixed point analysis. A fixed point is a point on the coordinate space which maps a function to the coordinate. For a two dimensional coordinate space, a fixed point of a function $f(x,y)$ is mathematically defined as $(x,y)$, see Ref.~\cite{11} such that-
\begin{eqnarray}\label{eqa6}
f(x,y)=(x,y).
\end{eqnarray}
From Brouwer's fixed point theorem it is known that {\it a 2D triangle $\Delta_2$ has a fixed point property}. This implies that any function which defines all the points inside a 2D triangle has a fixed point (for a detailed proof of this theorem refer to\cite{11}). Also the probabilities for choosing a strategy, represents points inside a square of side length 1. Thus
$S_{2,2}=(x,y)$ with $0<x<1$ and $0<y<1$, where $x$ represents the probability of choosing a strategy by row player and $y$ represents the probability of choosing a strategy by column player. It can shown that a triangle and a square are topologically equivalent\cite{11}, and this implies that if a triangle has a fixed point property, so does a square. So a function is constructed such that it represents all the points inside the square. To construct the function\cite{11}, a vector with coordinates ($u_1$, $u_2$) and another vector with coordinate ($v_1$, $v_2$) are defined as follows-
\begin{equation}\label{eqa9}
\left(\begin{array}{c} u_1\\u_2\end{array}\right)=A \left(\begin{array}{c} y\\1-y\end{array}\right),
\end{equation}
and
\begin{equation}\label{eqa10}
\left(\begin{array}{cc} v_1 & v_2\end{array}\right)= \left(\begin{array}{cc} x & 1-x\end{array}\right)B,
\end{equation}
where $x$ and $y$ are the probabilities to choose a particular strategy. $A$ and $B$ denote the respective payoff matrix for row player and column player. Using this, the fixed point function from Eq.~(\ref{eqa6}) is given by
\begin{equation}\label{eq44}
f(x,y)=\left(\frac{x+(u_1-u_2)^+}{1+|u_1-u_2|},\frac{y+(v_1-v_2)^+}{1+|v_1-v_2|}\right),
\end{equation}
where $(u_1-u_2)^+=\frac{u_1-u_2+|u_1-u_2|}{2}$ and $(v_1-v_2)^+=\frac{v_1-v_2+|v_1-v_2|}{2}$.
We determine $u_1$, $u_2$, $v_1$ and $v_2$ for the  payoff matrix as in Eq.~(\ref{eqa1}) and then the transformed one Eq.~(\ref{eqa2}). For the payoff matrix Eq. (\ref{eqa1}) we get the coordinates ($a_i's$ and $b_i's$ for i=1,2) as
\begin{eqnarray}\label{eqa3}
u_1=ay+b(1-y)\nonumber\\
u_2=cy+d(1-y)\nonumber\\
v_1=a'x+c'(1-x)\nonumber\\
v_2=b'x+d'(1-x).
\end{eqnarray}
Now for the transformed payoff matrix as in  Eq.~(\ref{eqa2}) the fixed point function is given by
\begin{equation}\label{eqa5}
f^t(x,y)=\left(\frac{x+(u_1^t-u_2^t)^+}{1+|u_1^t-u_2^t|},\frac{y+(v_1^t-v_2^t)^+}{1+|v_1^t-v_2^t|}\right).
\end{equation}
Again we determine the coordinates ($a_i's$ and $b_i's$ for i=1,2), as follows from Eq.~(\ref{eqa9},\ref{eqa10})
\begin{eqnarray}\label{eqa4}
u_1^t=(a+\lambda)y+(b+\mu)(1-y)\nonumber\\
u_2^t=(c+\lambda)y+(d+\mu)(1-y)\nonumber\\
v_1^t=(a'+\lambda')x+(c'+\mu')(1-x)\nonumber\\
v_2^t=(b'+\lambda')x+(d'+\mu')(1-x).
\end{eqnarray}
As we can see from Eq.~(\ref{eqa3}) and Eq.~(\ref{eqa4}), $u_1-u_2=u_1^t-u_2^t=(a-c)y+(b-d)(1-y)$ and $v_1-v_2=v_1^t-v_2^t=(a'-b')x+(c'-d')(1-x)$. Thus, $f^t(x,y)=f(x,y)$ which implies that the Nash equilibrium remains unchanged under the transformations as described before in Eq.~(\ref{eqa1}, \ref{eqa2}). 

\section{Acknowledgements }
CB thanks Science and Engineering Research Board (SERB) for funding under MATRICS grant ''Nash equilibrium versus Pareto optimality in N-Player games`` (MTR/2018/000070).

\section{Author contributions statement}
C.B. conceived the proposal,  S.S. did the calculations on the advice of C.B., C.B. and S.S. wrote the paper and analyzed the results. C.B reviewed the manuscript. 

\section{Additional information}
 \textbf{Competing interests} The authors declare no competing interests.


\begin{thebibliography}{99}
\bibitem{4}Game Theory in Action: An Introduction to Classical and Evolutionary Models, S. Schecter and H. Gintis, Princeton Univ. Press (2016).
\bibitem{5} C. Adami and A. Hintze, Thermodynamics of Evolutionary Games, Phys. Rev. E 97, 062136 (2018).
\bibitem{ai} Liviu Panait and Sean Luke,  Cooperative Multi-Agent Learning: The State of the Art. In Autonomous Agents and
Multi-Agent Systems. (Springer) 11(3) 387-434 (2005). 
\bibitem{12} M. A. Nowak, A. Sasaki, C. Taylor and D. Fudenberg, Emergence of cooperation and evolutionary stability in finite populations, Nature, 428, 646 (2004).
\bibitem{nowak} M. A. Nowak, Five Rules for the Evolution of Cooperation, Science (2006) 314, 1560-1563.
\bibitem{13} A. Bravetti, P. Padilla, An optimal strategy to solve the Prisoner’s Dilemma, Scientific Reports, 8, 1948 (2018). 
\bibitem{21} X. Chen and F. Fu, Imperfect vaccine and hysteresis, Proceedings of the royal society B 286(2019)
\bibitem{22} X. Chen and F. Fu, Social Learning of Prescribing Behavior Can Promote Population Optimum of Antibiotic Use, Frontiers in Physics 6, 139(2018)
\bibitem{19} G. Szabo and I. Borsos, Evolutionary potential games on lattices, Physics Reports, 624, 1 (2016).
\bibitem{1} S. Galam and B. Walliser, Ising model versus normal form game, Physica A 389, 481 (2010).
\bibitem{14} Sociophysics : An Introduction by P. Sen and B. K. Chakravarti, Oxford Univ. Press (2014).
\bibitem{15} K. Sznajd-Weron, J. Sznajd, Opinion evolution in closed community, Int. J. Mod. Phys. C 11 (6) (2000) 1157-1165. 
\bibitem{6} G. T. Landi, "Ising Model and Landau theory'' in  Lecture notes on undergraduate Statistical Mechanics (2017) available at: http://www.fmt.if.usp.br/~gtlandi/courses/stat-mech.html
\bibitem{11} Game Theory: A Playful Introduction, M. DeVos and D. A. Kent, American Mathematical Society (2016)
\bibitem{18} C. Benjamin and S. Sarkar, Triggers for cooperative behavior in the thermodynamic limit: a case study in Public goods game, Chaos 29, 053131 (2019).
\bibitem{10} S. Sarkar and C. Benjamin, Quantum Nash equilibrium in the thermodynamic limit, Quantum Information Processing 18:122 (2019). 
\bibitem{sar-benj-physa} S. Sarkar and C. Benjamin, Entanglement renders free riding redundant in the thermodynamic limit, Physica A 521, 607 (2019).

\end{thebibliography}
\end{document}